\documentclass[reprint,english,aps,pre,floatfix,showpacs,superscriptaddress]{revtex4-1}

\usepackage{amsmath}
\usepackage{graphicx}
\usepackage{amssymb}
\usepackage{babel}
\usepackage{color}
\usepackage{bm}
\usepackage{subfigure}

\renewcommand{\vec}[1]{\boldsymbol{#1} }
\makeatletter
\renewcommand{\thesubfigure}{\alph{subfigure}}   
\renewcommand{\@thesubfigure}{\thesubfigure)\hskip\subfiglabelskip}
\makeatother

\begin{document}

\title{Dynamic stratification in drying films of colloidal mixtures}

\author{Andrea Fortini}
\email{andrea.fortini@me.com}
\affiliation{Department of Physics, University of Surrey, Guildford GU2 7XH, United Kingdom}
\affiliation{Theoretische Physik II, Physikalisches Institut, Universit\"at Bayreuth, Universit\"atsstra{\ss}e 30, D-95447 Bayreuth, Germany}

\author{Ignacio Mart\'in-Fabiani}
\affiliation{Department of Physics, University of Surrey, Guildford GU2 7XH, United Kingdom}

\author{Jennifer Lesage De La Haye}
\affiliation{Laboratoire de Chimie, Catalyse, Polym\`eres et Proc\'ed\'es Universit\'e Claude Bernard Lyon 1, France}

\author{Pierre-Yves Dugas}
\affiliation{Laboratoire de Chimie, Catalyse, Polym\`eres et Proc\'ed\'es Universit\'e Claude Bernard Lyon 1, France}

\author{Muriel Lansalot}
\affiliation{Laboratoire de Chimie, Catalyse, Polym\`eres et Proc\'ed\'es Universit\'e Claude Bernard Lyon 1, France}

\author{Franck D'Agosto}
\affiliation{Laboratoire de Chimie, Catalyse, Polym\`eres et Proc\'ed\'es Universit\'e Claude Bernard Lyon 1, France}

\author{Elodie Bourgeat-Lami}
\affiliation{Laboratoire de Chimie, Catalyse, Polym\`eres et Proc\'ed\'es Universit\'e Claude Bernard Lyon 1, France}

\author{Joseph L. Keddie}
\affiliation{Department of Physics, University of Surrey, Guildford GU2 7XH, United Kingdom}

\author{Richard P. Sear}
\email{r.sear@surrey.ac.uk}
\affiliation{Department of Physics, University of Surrey, Guildford GU2 7XH, United Kingdom}


\begin{abstract}
In simulations and experiments, we study the drying of films containing mixtures of
large and small colloidal particles in water. During drying,  the mixture stratifies into a layer of the larger
particles at the bottom with a layer of the smaller particles on top. We developed a model to show that a gradient in osmotic pressure, which develops dynamically during drying, is responsible for the  segregation mechanism behind stratification.
\end{abstract}


\maketitle


Solid thin films on surfaces are often made by spreading a thin liquid film containing solid colloidal particles onto the surface, and allowing the liquid to evaporate~\cite{Routh:2013dm}. 
As we have known since the time of Robert Brown~\cite{brown}, colloidal
particles undergo Brownian motion; they diffuse.
As a  film dries, the water surface falls, pushing the colloidal
particles ahead of it. There is then competition between the particles'€™ Brownian
motion and the movement of the surface. The Brownian motion tends to distribute the colloidal particles
uniformly in the drying film, while the motion of the surface drives the system out of equilibrium.
It is known
that this competition determines the distribution of particles 
on the length-scale of the height $H$
of the film~\cite{Keddie:2010ta,Routh:2004jz,Trueman:2012er}.
Here we demonstrate a novel self-organization mechanism in colloidal mixtures, which occurs during solvent evaporation.  This mechanism separates large and small particles, to form a film
stratified by size.

To understand this mechanism, we start by considering
the simple case of a film containing only one species
of colloidal particle with a diffusion constant $D$.
The timescale
for diffusion across the height of the film is $H^2/D$.
During evaporation of the continuous solvent, the top surface moves down
with a velocity $v_{ev}$;  the evaporation timescale is $H/v_{ev}$.
The competition between these two timescales is quantified by the film formation P\'eclet number
Pe$_{film}=v_{ev}H/D$~\cite{Keddie:2010ta,Routh:2013dm}.
The drying film is near equilibrium if Pe$_{film}<1$, i.e., when the timescale for diffusion is smaller than
that for evaporation. In this case, evaporation only weakly
perturbs the vertical concentration profile, and the
profile remains almost uniform at all times. 
In the other limit, where Pe$_{film}>1$, diffusion cannot keep up with the moving interface, and particles accumulate near the descending interface at the top of the film \cite{gorce02,ekanayake09,Routh:2004jz,Keddie:2010ta,Routh:2013dm,Cardinal:2010be,Reyes:2007ku,Cheng:2013cp}. This description applies to one species of colloidal particle.

However, in paints and inks~\cite{Keddie:2010ta}, and often in nanofabrication~\cite{Rabani2003,Zhang:2010cn},
there are mixtures of different sizes (and types) of particles. Earlier
work has focused on the regime where the film formation P\'eclet number
of the large particles is greater than one, while that of the
smaller particles is less than one, and hence large
particles form the top layer
\cite{Luo:2008jd,Atmuri:2012ks,Trueman:2012er,Nikiforow:2010bi}.
There stratification is caused by the different rates at which small and large particles accumulate at the falling interface.

In this letter, we show that in the regime where both film P\'{e}clet numbers are much larger than one,  there is a generic tendency for the small particles to segregate in a layer on top of the larger particles.
This novel stratification mechanism is driven by a gradient of osmotic pressure and is found in both computer simulations and experiment on drying films containing mixtures of small and large colloidal particles.  
This is a previously unknown example of self-organization in a non-equilibrium process.

Moreover, this type of stratification is highly desirable because it allows  the independent control of the properties of the top and the bottom of a coating or self-organized nanostructure.  
The mechanism differs both from equilibrium phase separation
and the out-of-equilibrium Brazil-nut effect~\cite{Schroter:2006cc}.

\begin{figure}[tbh!]
\begin{center}
\includegraphics[width=8cm]{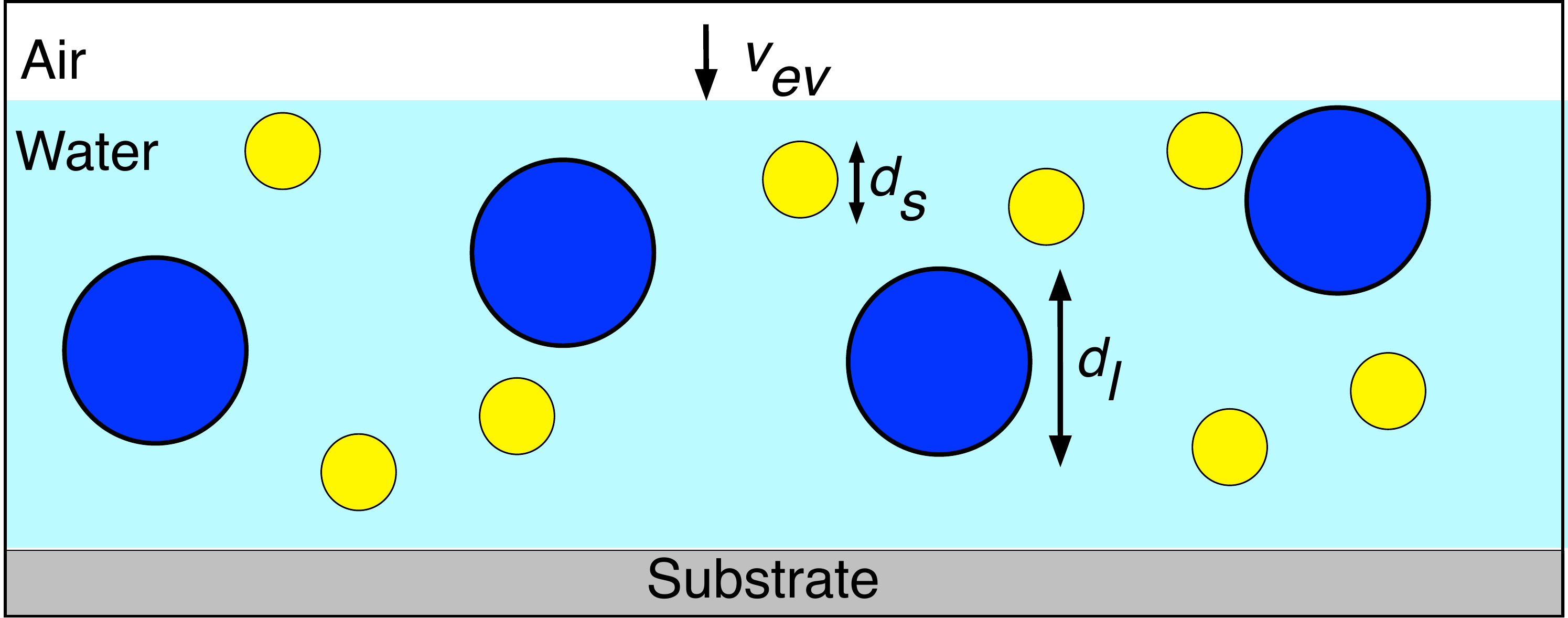}
\caption{Sketch of a wet film containing
a colloidal mixture of large particles with diameter $d_l$, and small particles with diameter $d_s$. The film
is bounded at the bottom by a substrate
and at the top by the air/water interface that falls with a velocity of $v_{ev}$. }
\label{sketch}
\end{center}
\end{figure}

 \begin{figure*}[htbp]
\begin{center}
\includegraphics[width=16cm]{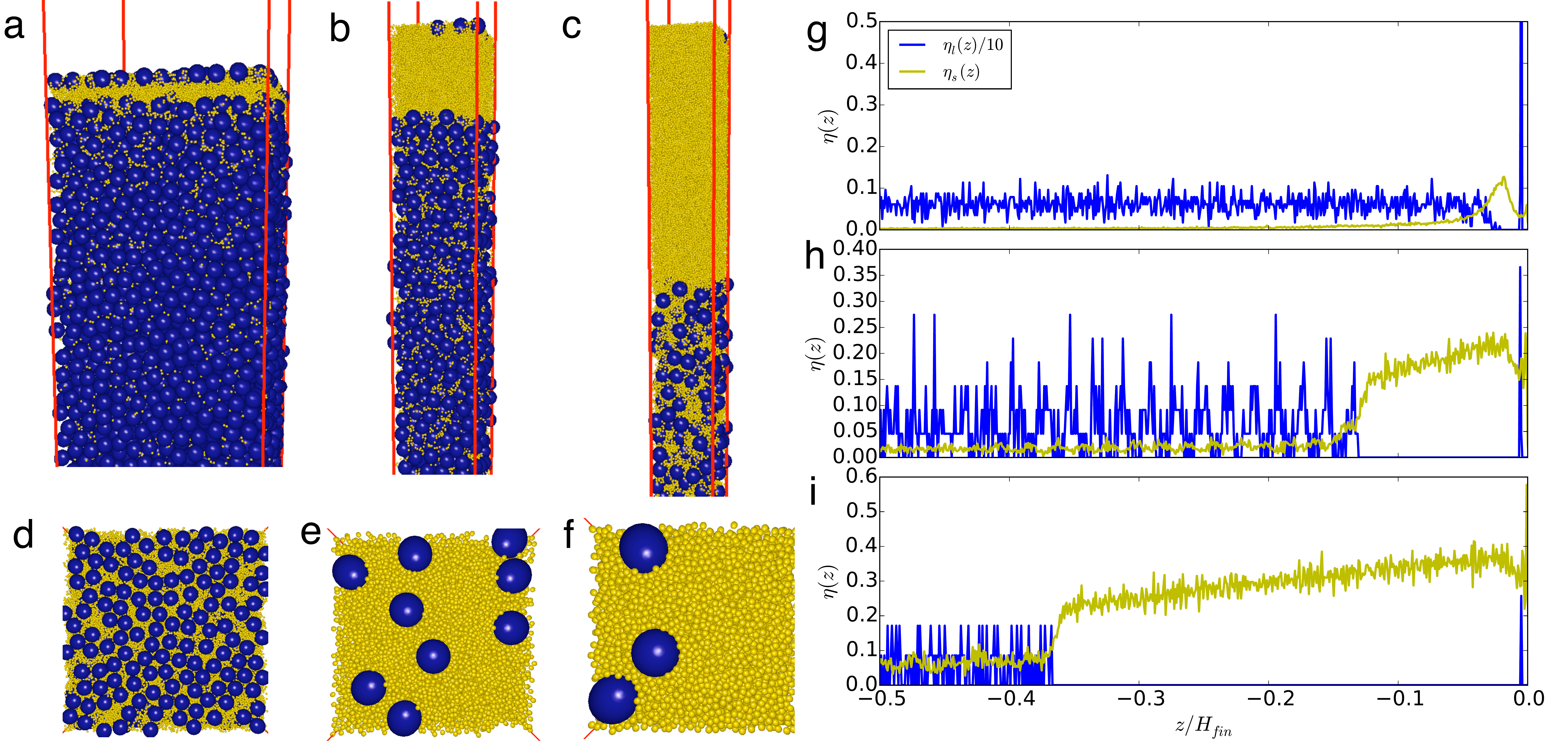}
\caption{(a)-(f) Simulation snapshots of the top third
of the simulation box, taken at the end of the run.
The small particles are shown in yellow (light gray),
the large particles in blue (dark gray), with size ratio $d_l/d_s$  of 7:1. 
In each case the system started at a total
volume fraction $\eta_0=0.1$ and height $H=1500~d_s$ and was run until
a final height of $H_{fin}=303~d_s$. The snapshots in (a), (b) and (c)
are for systems with increasing amounts of the small particles, the
number ratios are $N_r=5$, 29, and 151, respectively.
The snapshots in (d), (e) and (f) are the top views of the systems in (a), (b)
and (c), respectively. Due to the strong attraction between the surface and
the large particles, even in (c) and (f), where stratification is
strongest, we see some large particles
trapped at the surface. Their presence does not affect stratification.
(g)-(i) Volume fraction profiles of the small (yellow/light gray) and large (blue/dark gray) particles,
for $N_r=5$, 29 and 151, respectively. The bin width used in the profiles is 0.25$d_s$. 
}
\label{sim}
\end{center}
\end{figure*}

 \emph{Simulation}: Figure~\ref{sketch} illustrates the system under consideration; a two-component colloidal suspension of
large and small particles. The films are typically of the order of 
1000 particle diameters in height and macroscopic
in the other two directions. At the top is the water/air interface
and the substrate is at the bottom.

We carried out simulations on a binary mixture of spherical particles with diameters
$d_l$ and $d_s$; the size ratio $d_l/d_s=7$.
The  interaction between particles is that of screened charged particles, which is modeled by a short range repulsive Yukawa interaction. In contrast to the simulation of~\cite{Liao:2000ct}, our model assumes stable particles over the time scale of the evaporation.
The motion of the colloidal particles is simulated by Langevin dynamics~\cite{Plimpton:1995wl}, which includes Brownian diffusion
but neglects hydrodynamic flow. 
The simulation box has dimensions $L_x \times L_y \times H$.
To model a part of a large-area film that is far from
any edges, we apply
periodic boundary conditions in the $x$- and $y$-directions. 
Evaporation occurs along the vertical, $z$ direction. We model the air/water interface by a harmonic potential for the particles,
and evaporation is modeled by the potential's minimum moving
downward at a constant velocity $v_{ev}=0.05 \ d_s/\tau_B$, where $\tau_B=d_s^2/D_s$ is the Brownian time, and $D_s$ is the Stokes-Einstein diffusion coefficient of the small particles.
This gives a P\'{e}clet number for the small particles
of 75; that for large particles is $d_l/d_s=7$ times larger.
At the bottom we model the static substrate by the purely repulsive part of the Lennard-Jones
interaction.

In all cases we start with $N_s$ small and $N_l$ large particles,
such that the total initial volume fraction of the mixture $\eta_0=0.1$.
However, we varied the ratio $N_r=N_s/N_l$.
The system is equilibrated with a static top surface, and
after equilibration for a time 100 $\tau_B$, the downward movement of the model air/water interface begins. As the interface
moves downward at a constant velocity $v_{ev}$, both small and large particles
tend to accumulate below this moving interface.
As the interface moves down, an accumulation region forms and grows with time. This is a region where the density is higher and there is a density gradient.
A few large and small particles become trapped at the interface because of the effects of surface tension, while particles just below the surface diffuse normally. Inside the accumulation region where there is a density gradient, the large particles move away from the top region, creating a well-defined layer composed of only
small particles~\footnote{See  Fig.~\ref{s1} in Appendix}.
The width of the layer depleted of large particles grows in time, as is shown in Fig.~\ref{s1} of the Appendix. The layer continues to grow as long as the small particles can continuously filter through the large particles. This growth is hindered at higher volume fractions due to the slowing of the dynamics and the jamming of the small particles. The time for the segregation is larger than the time for evaporation for very high initial volume fractions.

The simulations are run until the accumulation front reaches the bottom substrate. The film height is initially $H=1500d_s$ and 
at the end of the simulation it is $H_{fin}<H$~\footnote{See Appendix for simulation details}. We do not examine the later stages of film formation.

In Fig.~\ref{sim}(a)-(f) we show snapshots of the top portion of the simulation box, taken when the accumulation
front has reached the bottom.
We see in Fig.~\ref{sim}(c) that
the small particles have formed a thick layer at the top
that has excluded the larger particles.
These larger particles have been pushed down into
a separate layer, with smaller particles in the interstitial spaces between the larger particles. 
The thickness of the layer of small particles at the top of the film is lower when the number of small particles is reduced,
as we can see in Fig.~\ref{sim}(a) and (b), but the layer is still present.
  
 {
Note that equilibrium mixtures of large and small hard spheres with a size ratio of 7:1 are completely miscible in the fluid phase \cite{dijkstra99}. At high volume fraction there is a broad region of coexistence between a crystal of the large particles and a fluid composed of mainly the small particles. But in our simulations stratification pre-empts crystallization, and
this stratificiation is inherently non-equilibrium
in nature; it is not due to an underlying equilibrium phase
separation.}

The stratification effect is general and occurs at different size ratios (Fig.~\ref{s2}) and for a range of initial volume fractions. With high volume fractions, we find a smaller width of the layer of small particles due to jamming effects of the small particle (Fig.~\ref{sdens}).

\begin{figure}[htbp]
\begin{center}
\includegraphics[width=8cm]{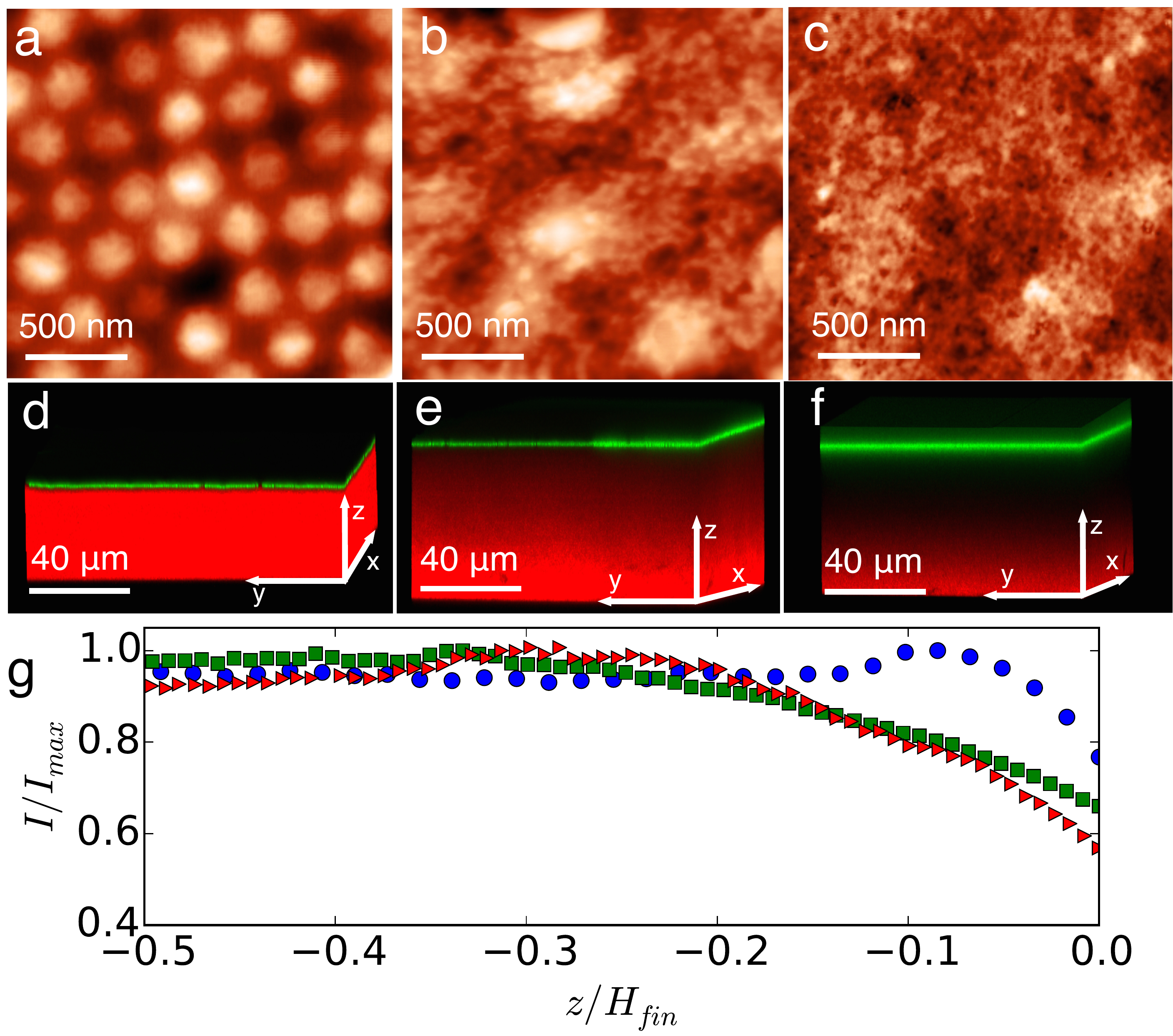}
\caption{Experimental results for dried films formed
of a binary mixture of colloidal particles of size ratio 7:1.
(a)-(c) Height AFM images for films with number
ratios $N_r=10$, 200 and 500,
respectively.
(d)-(f)   {Three dimensional confocal images} of the films for $N_r=10$, 200 and 500,
respectively. The large particles are labelled with a red dye,
while the small particles are unlabelled so the intensity of red
indicates the concentration of the larger particles.
To mark the position of the top surface, a drop of large (750 nm) green fluorescent particles was cast on the  dry film.
(g) Intensity of the red channel (large particles) as a distance from the top surface located at $z/H_{fin}$=0, for $N_r=10$ (blue circles), $N_r=200$ (green squares) and $N_r=500$ (red triangles). A correction was made for the depth dependence of the detected fluorescence intensity~\cite{Nikiforow:2010bi}. 
}
\label{exp}
\end{center}
\end{figure}
\emph{Experiments}:
Such a striking segregation has not been reported before, therefore we carried out an experimental investigation to confirm the findings.
Aqueous blends of colloidally stable acrylic copolymer particles ($d_l/d_s$ = 7)  {at total volume fraction $\eta_0=0.1$ and} varying number ratios were deposited on glass substrates with an initial wet thickness of approximately H =
700 $\mu$m.  We dried the samples at room temperature, leading to 
Pe$_{film}=$14 and 100, for the small and large particles, respectively. After film formation, the final films were characterized by means of atomic force microscopy (AFM) and scanning confocal microscopy. In order to visualize the population distribution of large (385 nm) particles within the sample, they were labelled with a red fluorescent dye (Rhodamine B); the small (55 nm) particles were unlabelled~\footnote{See Appendix for details of experiments and particle synthesis}.

Stratification is clearly seen in Fig.~\ref{exp},
although the layers
are less distinct than in our simulations.
Compare Fig.~\ref{sim}(g)-(i) to Fig.~\ref{exp}(g).
In our experiments, we see stratification for the
two mixtures with larger numbers of small particles,
$N_r=200$ and 500 in the confocal
images in Fig.~\ref{exp}(e) and (f), respectively.
The distribution is uniform with $N_r$=10. 

The surface coverage with small particles seen in the AFM images of Fig.~\ref{exp}(b) and (c) is also
consistent with stratification. The AFM images
show small particles at the top surface for the number
ratios $N_r=200$ and 500, but not for $N_r=10$.
Thus for mixtures of particles of size ratio 7:1, we see
stratification both in our computer simulations and experiments at sufficiently high number ratios.

\emph{Model:} In order to understand the segregation
of the large and small particles into layers,
we develop a physical model. 
In the evaporating film, density and hence pressure gradients
build up (See Fig.~\ref{s1}).
These gradients create forces that push particles
of all sizes down the gradients, and away from the surface.
Segregation results if these forces push the larger particles at
faster speeds than the smaller ones.
The speed of a particle
of diameter $d$ depends on the balance between the force $f(d)$ on
the particle, and the drag $\xi(d)$.

For simplicity, we will consider the case where a majority species
of diameter $d_m$ dominates the osmotic pressure, $P$,
but there is a trace amount of a species of a different diameter, $d_t$.

In the presence of a pressure gradient
$\partial P/\partial z$, the difference in pressure
between the top and bottom of a particle of diameter $d$
is $\approx d(\partial P/\partial z)$.
So the net downward
force on the particle
$f(d) \approx d^3 (\partial P/\partial z)$.

 {The friction coefficient of a particle of
diameter $d$ is $\xi(\eta,d)=  K(\eta,d) \xi_0$~\cite{VANDENBROECK:1981uj,Russel}.
Here $\xi_0=3 \pi d \nu$ is the
Stokes  friction coefficient, with $\nu$ the viscosity of water.}
$K(\eta,d)$ is the  sedimentation coefficient, defined as the ratio of the sedimentation velocity at volume fraction $\eta$ to that in
its dilute limit. 

At any point, the majority species will be pushed away from the interface
at speed $v(d_m)=f(d_m)/\xi(d_m)$. Segregation of the tracer 
particles is determined by their velocity relative to that
of the dominant species
\begin{equation}
\Delta v(d_t)=v(d_t)-v(d_m)=v(d_m)\left(
\frac{d_t^2K(\eta,d_m)}{d_m^2K(\eta,d_t)}-1\right) \ .
\end{equation}
At low densities  $K\simeq1$, therefore the downward
velocity of tracer particles relative to that of the majority
species varies as $(d_t/d_m)^2-1$, i.e., it increases quadratically
with the diameter of the tracer particles. Species larger than
the majority species move down faster than
the majority species, and segregation occurs with the
larger particles at the bottom. On the other hand, species smaller than the majority species move down slower than
the majority species, resulting in smaller particles accumulating at the top. 
The functional form of $K(\eta,d)$ has been the focus of many studies in theory, simulations and experiments~\cite{Hayakawa:1995vc,Russel,Bowen:2000jf} and  does not depend on P\'{e}clet number~\cite{Padding:792074}. 
For Brownian particles, the diameter dependence at high density is  {$ K \simeq d$}, which leads to a segregation velocity that scales as $(d_t/d_m)-1$, i.e., it would still lead to stratification.  {However, at high volume fraction the dynamics of the system slows considerably and the time scale for the segregation mechanism could become larger than the time for solvent evaporation~\cite{Peppin:2006uq}. }

\emph{Test of the model:} Our simple theoretical model makes a striking prediction:
larger particles move down relative to the majority species, while smaller ones
move up.
This can be independently verified by simulating mixtures
with a majority species plus both smaller and larger particles.
Therefore, we simulated a ternary mixture of particles:
a majority species with  diameter  $d_m$, and two minority components with size ratios $d_s/d_m$=0.8 and $d_b/d_m$=1.2. The initial average volume fractions are  $\eta_m=0.05$, $\eta_s=0.003$, $\eta_b=0.0052$ and the evaporation velocity is $v_{ev}  = 0.1 \ d_m / \tau_B$.

\begin{figure}[htbp]
\begin{center}
\includegraphics[width=8cm]{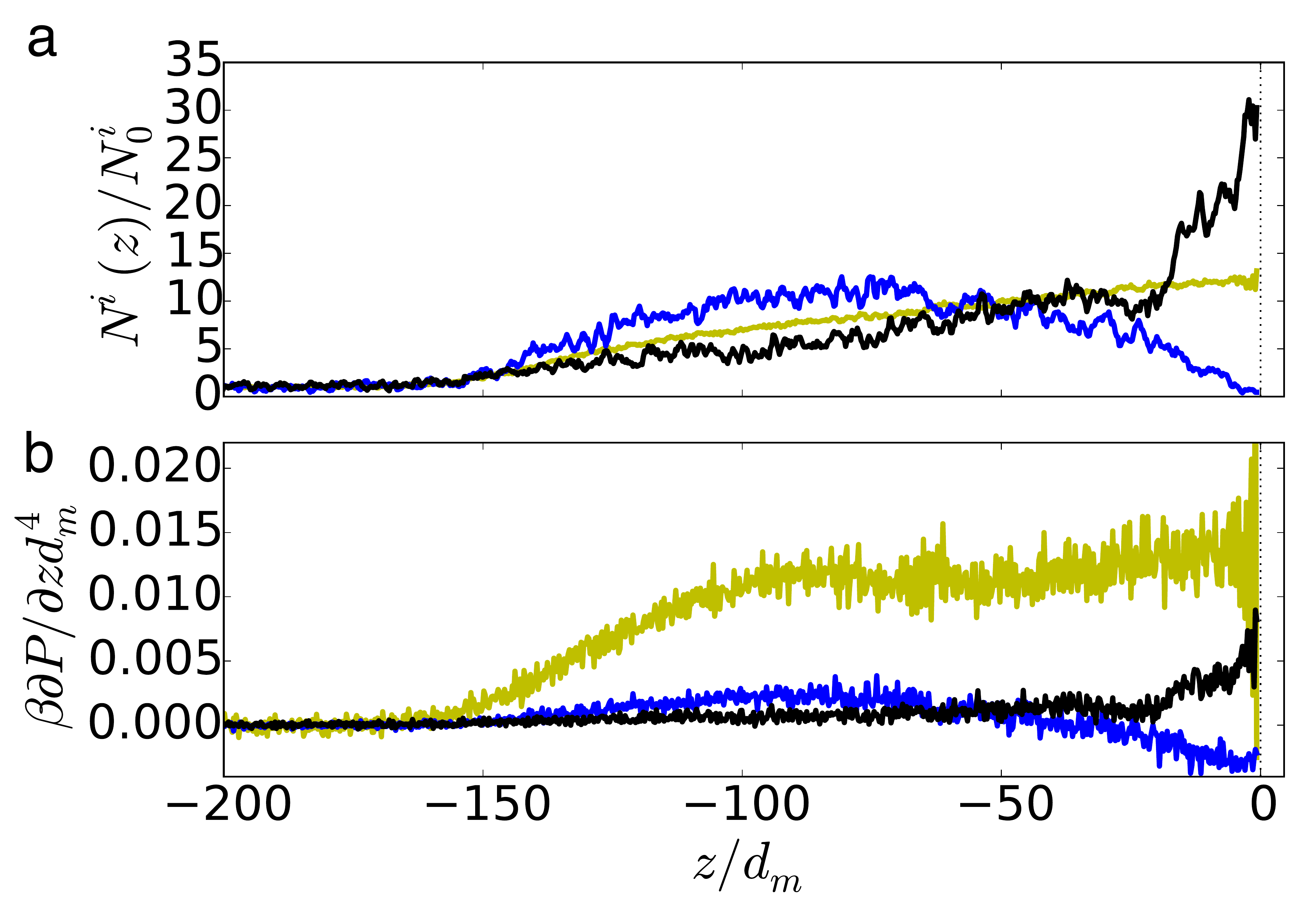}
\caption{Density and pressure gradient 
profiles in the drying film at time t=4350~$\tau_B$ for ternary mixtures. The top surface is at $z/d_m=0$.
The yellow (light gray) curves represent the majority species, the blue (dark gray) curves represent the larger species,
$d_b/d_m=1.2$, and the black curves represent the smaller species, $d_s/d_m=0.8$.
(a) Densities of particles as a function of the distance from the
interface, plotted as $N^i(z)/N^i_0$, where
$N^i(z)$ is the density of species $i=m,s,b$ and $N^i_0$
is the initial number density.
(b)  Vertical gradients of the osmotic pressure as a function of the distance from the top interface.
The component of the dominant species makes the greatest pressure  contribution. The vertical dotted lines indicate the position of the top interface.
}
\label{p3}
\end{center}
\end{figure}

The results for the ternary mixture are shown in Fig.~\ref{p3}. 
In Fig.~\ref{p3}(a) we see that the moving surface has created
gradients in the density and hence in the pressure of
the majority species (shown in yellow/light gray), of width $\approx 150~d_m$.
The gradient of the osmotic pressure is plotted in
Fig.~\ref{p3}(b).
The dominant force is due to the osmotic pressure gradient of the majority species. 

The larger species are on average farther from the top
surface than the majority species. Note the maximum
in their density (shown in blue/dark gray) around $100 \ d_m$ below
the surface. By contrast, the smaller particles
(shown in black) are accumulating near the top surface.
As predicted, at our large film formation P\'{e}clet numbers,
mixtures of particles of different sizes are unstable
with respect to stratification into layers, with the smallest
particles at the top, and the largest at the bottom.

In our simulations we neglected any effect due to hydrodynamic
flow of the solvent. Modeling flow for our systems
of many thousands of particles is not computationally feasible.
Flow is present in the experiments of course.
When the volume fraction of the drying suspension is changing, there will
be relative motion of the particles and water, which will generate forces on the particles acting toward
the surface. 
On one hand, these forces will push larger particles toward the surface and counteract the segregation of small particles. 
On the other hand, the majority species will be pushed toward the surface and create larger osmotic pressure gradients that will enhance segregation. 
We cannot calculate these forces, but we note that the effect we describe
here is very robust in the simulations. 
Furthermore, we see the effect in experiments where there is hydrodynamic flow
of water. Hence, we believe that stratification does
occur in the presence of forces due to hydrodynamic flow.

\emph{Discussion and Conclusion:}
In both computer simulations and experiments on drying colloidal mixtures,
we found stratification. The smaller particles excluded
the larger particles and formed a layer at the top of the drying film.
This is a purely out-of-equilibrium effect; it is driven by the moving interface.
The moving interface causes a density,
and hence a pressure, gradient in the drying film, and this pressure gradient
pushes larger particles away from the moving interface faster than it pushes smaller
particles. We developed a physical model for this process, and the model correctly predicted
the behavior of both small and large particles.

Diverse technologies, including inkjet printing~\cite{Tekin:2008bu}, coatings on pharmaceutical tablets~\cite{Obara:1995wl,Lecomte:2004jo}, agricultural treatments on crops~\cite{Faers:2008bc,Taylor:2011fi,Hampton:2012kf}, synthetic latex paints, adhesives~\cite{Keddie:2010ta}, and cosmetics (such as sun screen~\cite{Wissing:2001ut}), rely on films derived from mixtures of colloidal particles. Our discovered mechanism will be useful whenever the properties of the top and the bottom of a coating need to be controlled independently via a one-step deposition process.

\appendix
\section{Evolution of density profiles}

\begin{figure*}[htbp]
\begin{center}
\includegraphics[width=13cm]{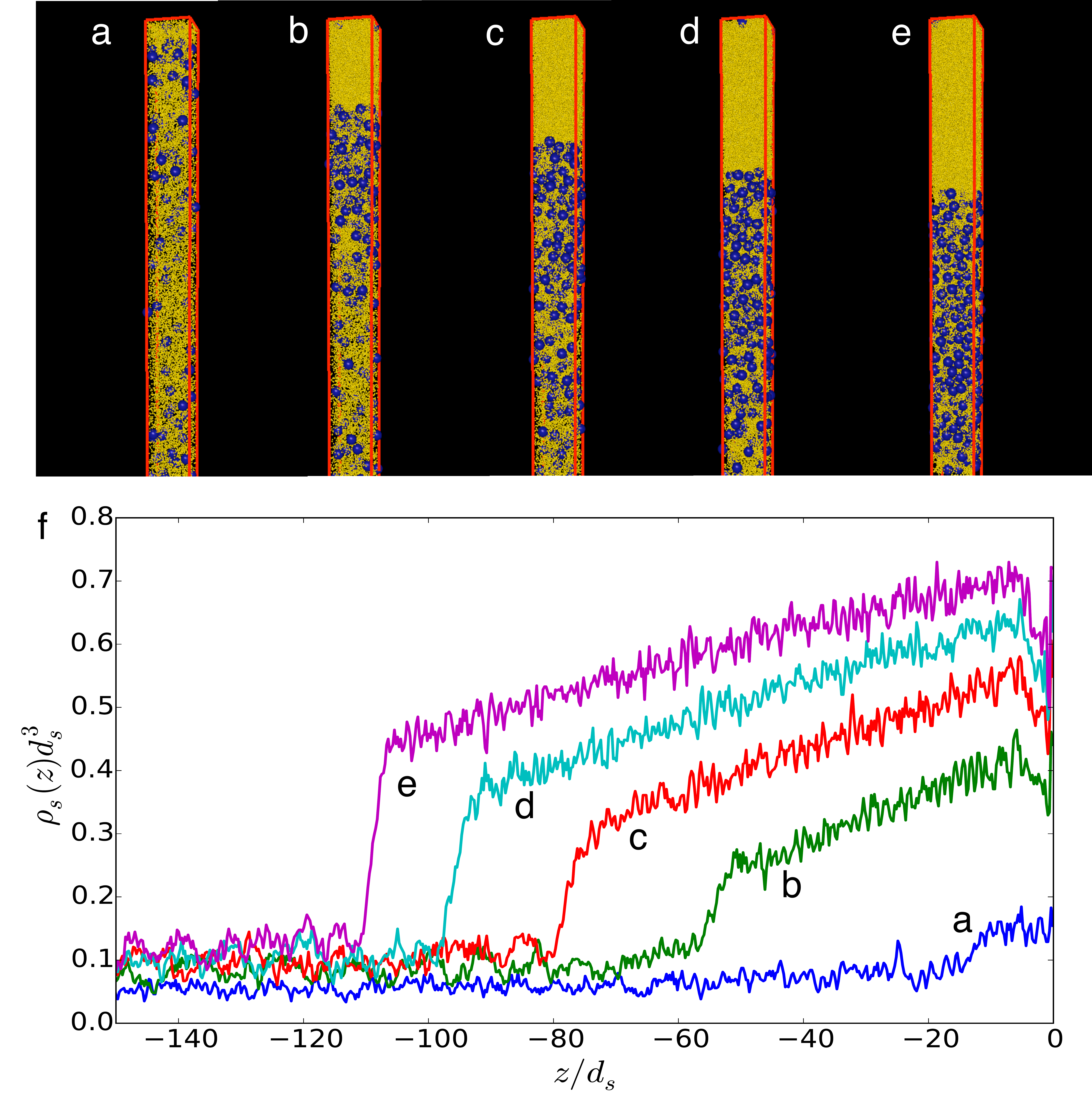}
\caption{Evolution with time of the density profiles of the small particles, for a mixture with size ratio $d_l/d_s=7$
and number ratio $N_r=151$.
(a)-(e) are snapshots at times
$t=7.5 \times 10^2 \tau_B$, $6.6 \times 10^3 \tau_B$,
$1.25 \times 10^4 \tau_B$, $1.83 \times 10^4 \tau_B$
and $2.4 \times 10^4 \tau_B$, respectively.
The corresponding density profiles are plotted in (f);
$\rho_s$ is the number density of small particles.
The top surface is at $z/d_s=0$ and is at the right.
}
\label{s1}
\end{center}
\end{figure*}

The evolution of the stratified structure during evaporation and of the related density profiles is shown in Fig.~\ref{s1}.

\section{Stratification at different size ratios and initial volume fractions}

\begin{figure}[htbp]
\begin{center}
\includegraphics[width=8cm]{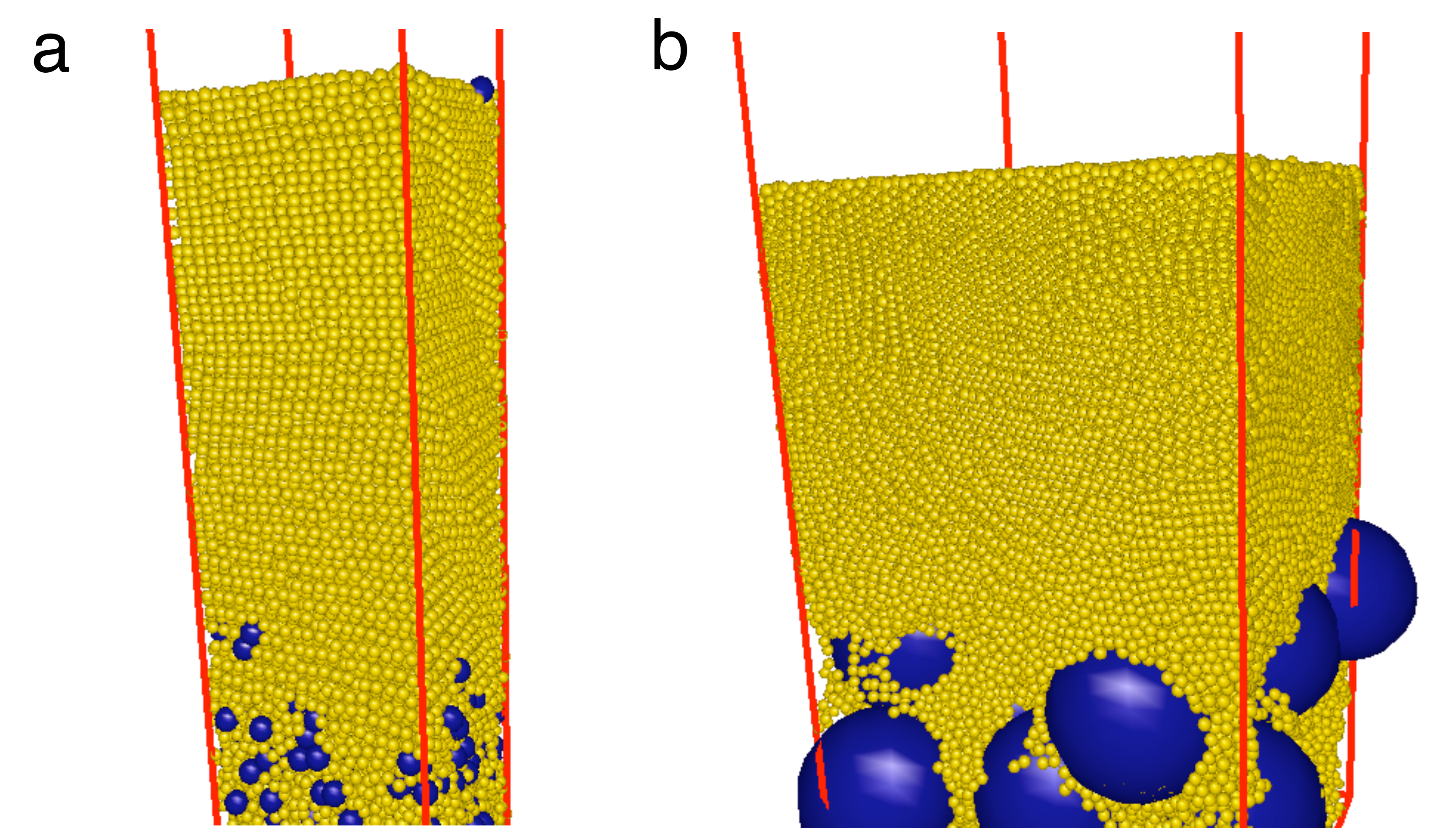}
\caption{Snapshots of the top regions
of the final configurations.
(a) Size ratio $d_l/d_s$=2, and $N_r$=17. (b) Size ratio $d_l/d_s$=14, and $N_r$=9000. }
\label{s2}
\end{center}

\end{figure}

Snapshots of the final configuration obtained in simulations of binary mixtures with size ratio $d_l/d_s$=2, and 14 are shown in Figure~\ref{s2}. The formation of a top layer depleted of large particles is clearly visible in both cases. 

\begin{figure}[htbp]
\begin{center}
\includegraphics[width=8cm]{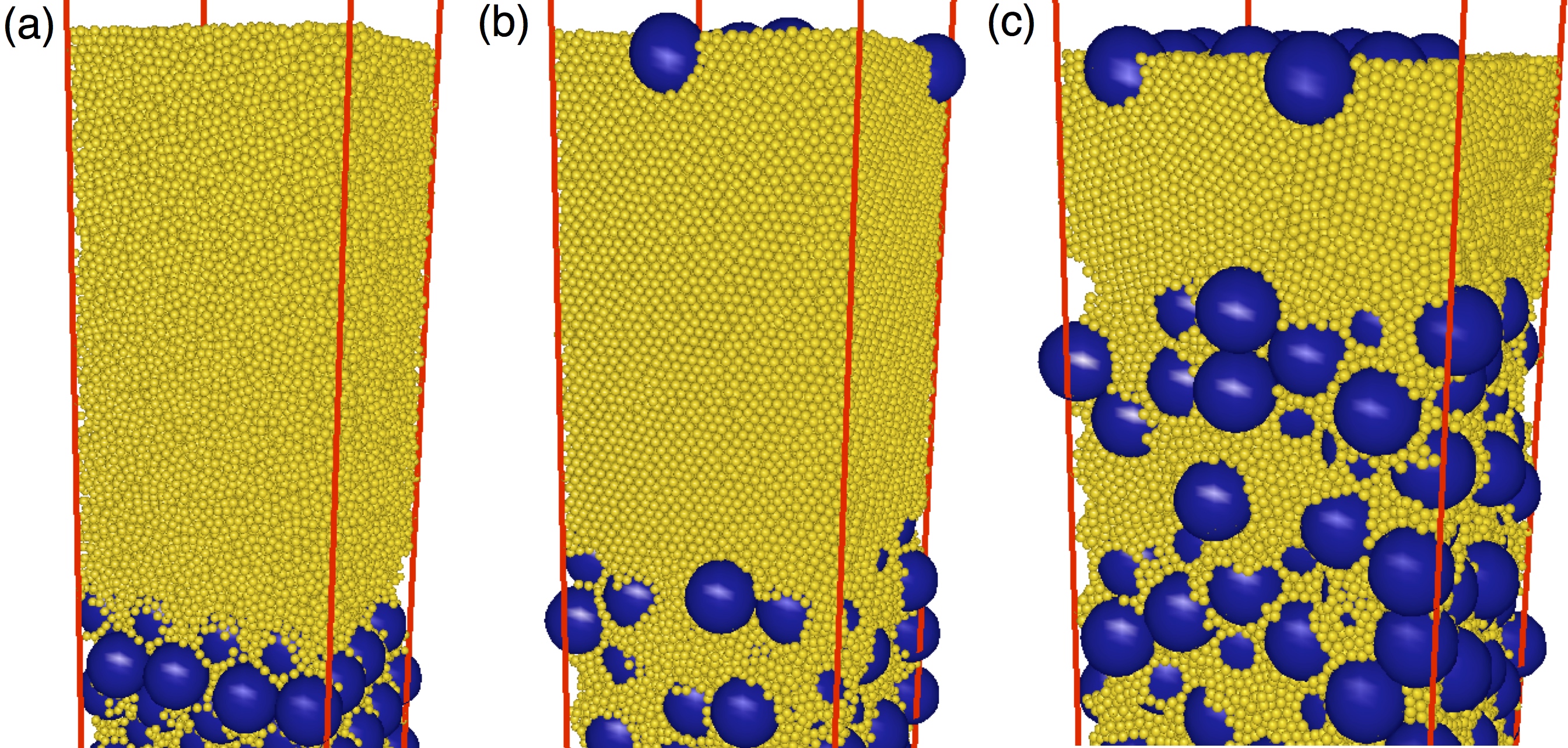}
\caption{Snapshots of the top regions 
of the final configurations for mixtures with  size ratio 7:1 and number ratio $N_r$=150 for different initial volume fractions.
(a) $\eta_{0}$=0.1 (b) $\eta_{0}$=0.2 (c) $\eta_{0}$=0.4.}
\label{sdens}
\end{center}

\end{figure}

In Fig.~\ref{sdens} we show snapshots of the top regions  of the final configurations for initial volume fractions $\eta_{0}$=0.1 (a) $\eta_{0}$=0.2 (b), and $\eta_{0}$=0.4 (c), obtained with an evaporation velocity $v_{ev}=0.1$.
The layer with only small particles
is visible in the final film layer regardless of the initial volume fraction, but as the initial particle density increase, the segregation mechanism becomes less efficient.

\section{Details of the Simulation Method}

We run computer simulations of mixtures
of spherical particles in a simulation box with dimensions $L_x \times L_y \times H$. Periodic boundary conditions are used in the $x$- and $y$-direction, while in the $z$-direction the box is delimited at the bottom by a hard substrate, and at the top by a soft wall, which models an air-water interface.  
We do not explicitly simulate the solvent water molecules but describe the motion of the colloidal particles using  Langevin dynamics, which includes Brownian diffusion effects and neglects hydrodynamic flow.  The equation of motion for a particle $i$ with diameter $d_i$, mass $m_i$ at position ${\vec r}_i$ is 
\begin{equation}
m_i  \ddot {\vec r}_i= - \sum_{i<j} \nabla U_{ij}(\vec r_i,\vec r_j) - \xi_i  \dot {\vec r}_i + \delta F_i \ ,
\label{le}
\end{equation}
where  $\xi_i=3 \pi \nu d_i$ is the friction coefficient and $\nu$ is the viscosity of the solvent. The term $\delta F_i$ is a random force  sampled from a Gaussian distribution with width  $\sqrt{k_B T \xi_i}$, where $k_B$ is the Boltzmann constant and $T$ the temperature. The Langevin dynamics was carried out using the LAMMPS package~\cite{Plimpton:1995wl} with a time step  dt = 0.001 $\tau^s_B$ = 0.0025 $t_0$, where $\tau^s_B=d_s^2/D^s_0$  is the Brownian time of the small particles, and $t_0$ is the standard Lennard-Jones unit of time in LAMMPS. We also used a friction parameter $ \xi_i=100 \ d_i$ and a temperature of 40, both in standard Lennard-Jones LAMMPS units. 

The  interaction energy models screened charged particles, i.e.,  the interparticle potential energy $u_{ij}$ between particles $i$ and $j$, either small or large,  of diameters $d_i$ and $d_j$ is a short range Yukawa interaction
\begin{equation}
u_{ij}(r)/k_BT=
\begin{cases} 
 \frac{\epsilon}{k_B T} \displaystyle e^{-\kappa(r-\sigma)} &  r < r_c\\
0 & r \ge r_c
\end{cases}
\end{equation} 
where $r$ is the centre-to-centre particle distance, $\epsilon$ is the contact energy, and $\kappa$ determines the steepness of the potential. These two parameters influence the overall softness of the potential. In this simulation, we choose $ \epsilon/k_B T=25$ and  $\kappa d_s$=20.
The distance  $\sigma=(d_i+d_j)/2$ and the cut-off is $r_c= (d_i+d_j)/2 +d_s$. 

The binary mixture is modeled by  small particles with diameter $d_s=1$ and mass $m_s=1$ and large particles with diameter $d_l=7 d_s$ and mass $m_l=d_l^3 m_s$. 
Because of the soft interaction between particles, it is possible to define an effective diameter of the spheres using the Barker-Henderson relation~\cite{Barker1976}
 \begin{equation}
 d^{\rm eff}=d+\int_d^\infty (1-\exp[-u_{ij}(r)/k_BT]) dr \ ,
 \end{equation}
which gives the distance where the repulsive interaction is of the order of 1 $k_B T$. For our parameters, the  effective diameters are $ d^{\rm eff}_l=7.19$ and $ d^{\rm eff}_s=1.19$, for large and small particles, respectively. The effective size ratio therefore is  $ d^{\rm eff}_l/ d^{\rm eff}_s$=6.04.
The model neglects deformability and coalescence of the particles, which can occur when film forming particles are used in experiments.

The interaction  $U_{iw}(z)$ between  particle $i$ with diameter $d_i$ and the hard  substrate at the bottom is modeled by 
\begin{equation}
U_{iw}(h)/k_B T=
\begin{cases} 
 \frac{\epsilon _w}{k_B T} \displaystyle \left ( \frac{d_i}{h} \right )^{12} &  h < d_i/2\\
0 & h \ge z_c
\end{cases}
\end{equation} 
where $\epsilon_w/k_B T=100$ determines the strength of the repulsive interaction and $h$ is the distance of a particle from the substrate. 

We model the solvent evaporation process by a moving soft wall, which pushes the particles toward the bottom substrate at constant velocity $v_{ev}$. The position of the soft-wall (interface) as a function of time is defined by $z_{int}(t)=H-v_{ev} t$.
The interaction between the soft wall and a particle $i$ with diameter $d_i$ is described by a harmonic potential, which models the Pickering effect due to the change in interfacial free energy when particles are trapped at a interface~\cite{Pieranski1980} 
\begin{equation}
U_i(z)/k_B T=\frac{\alpha_i}{k_B T} (z-r_0-z_{int}(t)) \ ,
\end{equation}
where $z$ is the particle coordinate, and $r_0$ determines the contact angle $\theta= \arccos(2 r_0/d_i)$. We have chosen $r_0=d_l/4, d_s/2$, for large and small particles, respectively.  The strength of the air-water interface attraction was chosen to be proportional to the area of the particle, i.e., $\alpha_i/k_B T=1000 \ (d_i/d_s)^2$. 
Effects like the capillary attraction between the particles trapped at the interface or effective dipolar interactions are neglected in this model.

\section{Details of the Experiment}

We  investigated  the room temperature drying of latex particle mixtures experimentally. The initial colloidal dispersions were prepared by  blending two acrylic copolymer latices with different mean particle sizes. Both types of particles were mutually repulsive and were colloidally-stable in water initially and when mixed together.
 
The large particles were made of a copolymer of methyl methacrylate and $n$-butyl acrylate in a weight ratio of 40/60. The initial solids content is 20 wt\%. They were synthesized by radical emulsion polymerization using Synperonic® NP30 and sodium dodecylsulfate surfactants (99/1 wt\% ratio) at a total concentration of 3 g L$^{-1}$. Sodium persulfate (0.5 wt\% relative to monomers) was used as the initiator. 

The z-average diameter was determined by dynamic light scattering (DLS) (NanoZS from Malvern Instruments) to  $D_z$=385 nm. These particles were labelled with fluorescent Rhodamine B (0.2 wt\% based on monomers). Analysis of the supernatant after centrifugation found that there was no Rhodamine B in the aqueous phase. The particles were stabilized by initiator fragments and a combination of non-ionic and anionic surfactants (with a weight ratio of 99/1).  
 
The small particles are composed of amphiphilic block copolymers obtained by polymerization-induced self-assembly (PISA)~\cite{ref1}. Controlled radical polymerization (namely, reversible addition-fragmentation chain transfer (RAFT) polymerization) of methacrylic acid (MAA) was performed to obtain firstly a PMAA macroRAFT agent (about 4000 g mol$^{-1}$)~\cite{ref2}, which was then chain-extended with a mixture of n-butyl acrylate (BA) and styrene (S) (55/45 wt\%) to form self-stabilized particles~\cite{ref3}. A z-average diameter $D_z$=55 nm with a dispersity of 0.064 were measured by DLS. Electrostatic stabilization was provided by anionic sulfate groups contained in the initiator. The pH of the colloidal mixture was measured to be 3.5. At this pH, which is below the value of PMAA's pKa of 5.5, the PMAA chains at the particle surface are only weakly ionized and are collapsed~\cite{wang1}. Hence, the PMAA is expected to offer neither significant steric stabilization nor charge stabilization.

After blending the calculated amounts of these two dispersions to achieve the desired number ratio, we added deionized water in order to match the initial solids content used in the simulations. The final solids content was always in the range of 9-13 wt.\% for all dispersions.
Films of these blends were cast on glass substrates ($18\times 18$ mm$^2$), previously cleaned with acetone and a UV ozone treatment (Bioforce Nanosciences, model UV.TC.EU.003).
 
Height and phase images of the top surface of the films were acquired by atomic force microscopy (AFM), using an NT-MDT Ntegra Prima microscope with intermittent contact. Images were analyzed using  NOVA software.
 
A Zeiss LSM510 confocal microscope (on an Axiovert 200M microscope) was used to obtain stacks of plane images at different depths within the sample. A drop of large (750 nm diameter, purchased from Fluoresbrite) green fluorescent particles was cast on top of the dried films.
They provided a marker for the top surface position. The green and red fluorochromes were excited using an argon laser (488 nm) and a HeNe laser (543 nm), respectively.
 Two-dimensional images $132\times 132~\mu$m$^2$) were acquired every 0.5 $\mu$m when moving from the substrate at the bottom toward the top of the dry film. Results were analysed using the image processing package Fiji (a version of Image J). The position of the fluorescent green marker particles in the images was used to define the top surface. A second-order polynomial equation was fit to the intensity of the red channel as a function of depth from the surface and then used to define a baseline, to correct for the depth-dependence of the detected intensity. The corrected intensity was normalized by dividing by the maximum intensity in the profile, and the vertical position was normalized by dividing by the film thickness.

\begin{acknowledgments}
We acknowledge funding from the European Union Seventh Framework Programme BARRIER-PLUS project
(FP7-SME-2012-2, no.~304758).
\end{acknowledgments}

\bibliography{refs}

\end{document}